\documentclass[letterpaper,prl,twocolumn,superscriptaddress,preprintnumbers]{revtex4}
\usepackage{bm,graphicx}
\usepackage{color}

\begin{document}
%\preprint{\textsf{\textbf{\color{blue} To appear in Phys. Rev. B (Rapid Comm)}}}

\title{Non-collinear Exchange Coupling in Trilayer Magnetic Junction and its Connection to Fermi Surface Topology}
\author{Wen-Min Huang}
\affiliation{Department of Physics, National Tsing-Hua University, Hsinchu 300, Taiwan}
\author{Cheng-Hung Chang}
\affiliation{Institute of Physics, National Chiao-Tung University, Hsinchu 300, Taiwan}
\affiliation{Physics Division, National Center for Theoretical Sciences, Hsinchu 300, Taiwan}
\author{Hsiu-Hau Lin}
\affiliation{Department of Physics, National Tsing-Hua University, Hsinchu 300, Taiwan}
\affiliation{Physics Division, National Center for Theoretical Sciences, Hsinchu 300, Taiwan}
\date{\today}
\begin{abstract}
We investigate the non-collinear exchange coupling across the trilayer magnetic junction composed of an intermediate layer with Rashba interaction and two sandwiching ferromagnetic ones. To compute the mediated exchange coupling, one needs to go beyond the single-particle argument and integrate over the contributions from the whole Fermi surface. Surprisingly, we find that the topology of the Fermi surface plays a crucial role in determining whether the oscillatory RKKY or the spiral interactions would dominate. At the end, we discuss the connection of our numerical results to experiments and potential applications.
\end{abstract}
%\pacs{}
\maketitle

The central theme of spintronics is to manipulate the extra spin degrees of freedom in condensed matter systems\cite{Wolf01,Zutic04,MacDonald05,Sun04b,Sun06}, as compared with the traditional electronic devices where only the charge part was utilized. One of the classic examples, which merges charge and spin sectors together in a single device is the spin field effect transistor (SFET) proposed by Datta and Das\cite{Datta89} more than a
decade ago. It was suggested that the Rashba interaction, whose strength is controlled by the gate voltage, causes the spins of the itinerant carriers to spiral and can be used to modulate the transport currents. Another more recent proposal, now under the name of spin Hall effect\cite{Murakami03,Sinova04}, explores the possibility to generate spin currents (or spin accumulations) by electric gates via spin-orbital interactions and has received some primitive verifications in experiments.

\begin{figure}
\centering
\includegraphics[width=7cm]{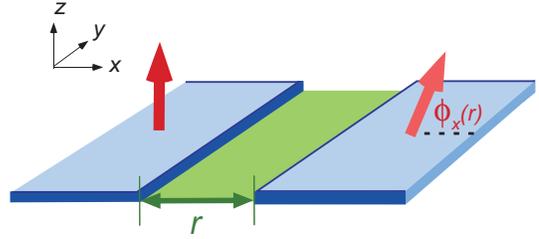}
\caption{\label{TMJ}
(Color online) Schematic plot for the trilayer magnetic junction where $\phi_x(r)$ is the non-collinear angle between the ferromagnets.
}
\end{figure}

On the other hand, it is less explored how the spin-orbital interaction will reshape our understanding in the more conventional magnetic junctions\cite{Sun04a}. In this Letter, we study the non-collinear exchange coupling across the F/N/F trilayer magnetic junction (TMJ) as shown in Fig.~\ref{TMJ}, where the intermediate layer consists of semiconductors (such as GaAs) with significant Rashba interaction. Since the TMJ resembles the Datta-Das SFET, one may naively guess that the carrier-mediated exchange coupling can be explained following similar arguments. Within the single-particle picture, Datta and Das had demonstrated the spins of the itinerant carriers undergo precessions due to the Rashba interaction and, therefore, lead to the non-collinear coupling between the ferromagnetic layers. 

However, there is a sharp difference between SFET and TMJ. In the SFET, the time-reversal symmetry is broken (by the source-drain bais which drives the current) while, in the TMJ, the symmetry is preserved and gives rise to Kramers degeneracy. In fact, this non-trivial degeneracy has a profound influence upon the effective exchange coupling across the junction. It turns out that the single-particle picture (with specific momentum) employed in Datta-Das' original work\cite{Datta89} fails to explain the magnetic behavior because the inclusion of the {\em whole} Fermi surface is crucially important. For instance, the quantum interferences between the Kramers-degenerate patches of the Fermi surfaces give rise to the oscillatory RKKY interaction. To explore the subtle competition between the spiral and the RKKY interactions, one needs to integrate over the whole Fermi surfaces by numerical approach.

\begin{figure}
\centering
\includegraphics[width=7.5cm]{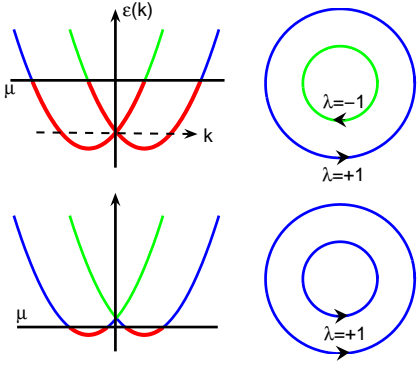}
\caption{\label{topology}
(Color online) Band structure of the Rashba Hamiltonian and the Fermi surface topology. In weak Rashba regime, the Fermi surface consists of two concentric circles with opposite chiralities while there is only one chirality present in strong Rashba regime.
}
\end{figure}

The outcome is rather surprising -- the dominance of either spiral or RKKY interactions depends on the topology of the Fermi surface. When the Rashba interaction is weak (compared with the Fermi energy), the Fermi surface consists of two cocentered circles with opposite chiralities. In this regime, the RKKY interaction dominates over the spiral. However, as one gradually increases the strength of the Rashba interaction, the inner Fermi circle shrinks to zero first, then reappears again but with the same chirality as the outer circle (as shown in Fig.~\ref{topology}). The topological change of the Fermi surface magically alters the dominant coupling from the of RKKY to the spiral. Therefore, in the strong Rashba limit, the non-collinear exchange coupling mainly comes from the spiral interaction with minor quantum corrections. It is rather amazing that the transition between different magnetic behaviors coincides with the change of the Fermi surface topology.

In the following, we present the analytic arguments and numerical results which support the claims we made in above. First of all, we model the intermediate layer by 2D electron gas with Rashba interaction,
\begin{equation}\label{TH}
H = \int d^2r\: \Psi^{\dag} \left[\frac{k^2}{2m^*}\textbf{1}
+\gamma_{R} (k_y\sigma^x- k_x\sigma^y)\right] \Psi,
\end{equation}
where $\gamma_{R}$ is the strength of the Rashba interaction and 
$\Psi^{\dag}, \Psi$ are the two-component spinors of the creation/annhilation operators for itinerant carriers. The Rashba Hamiltonian can be brought into its eigenbasis in momentum space,
\begin{equation}
\varphi_{k\lambda}(\vec{r}) = e^{i\vec{k}\cdot\vec{r}} u_{\lambda}(\phi)
= \frac{e^{i\vec{k}\cdot\vec{r}}}{\sqrt{2}}\left(
\begin{array}{c}
-i\lambda e^{-i\theta_{k}}\\
1 
\end{array}\right)
\end{equation}
with $\theta_{k} = \tan^{-1} (k_{y}/k_{x})$. Due to the spin-orbital interaction, spin is no longer the good quantum number but replaced by the chirality instead,
\begin{eqnarray}
\lambda = (\hat{\bm{k}} \times \hat{\bm{s}})\cdot \hat{\bm{z}} = \pm 1,
\end{eqnarray}
where the hat denotes the unit vector. It is important to remind the readers that, under the time reversal transformation, both momentum and spin reverse their directions and make the chirality invariant.

After integrating out the itinerant carriers\cite{Sun04a}, the exchange coupling between the ferromagnetic layers is described by an effective Heisenberg Hamiltonian, $H_{\rm eff} = \sum_{ij} J_{ij} S^{i}_{L}S^{j}_{R}$. Within the linear response theory, the mediated exchange coupling is proportionally to the spin susceptibility tensor,
\begin{equation}\label{sus1}
\chi_{ij}(\textbf{r})=\int_{0}^{\infty}dt\left\langle\left\langle i
\left[\sigma^{i}(\vec{r},t),\sigma^{j}(0,0)\right]\right\rangle\right\rangle
e^{-\eta t}.
\end{equation}
Here $\sigma^{i}(\vec{r},t)=\sum_{\alpha\beta}\psi^{\dag}_{\alpha}
(\vec{r},t)\sigma^{i}_{\alpha\beta}\psi_{\beta}(\vec{r},t)$ is the spin density operator for the itinerant carriers. Besides, $\langle\langle...\rangle\rangle \equiv \textrm{tr} [e^{-\beta H} ...]$ represents the thermal average at finite temperature and $\eta$ is the spin relaxation rate. Transforming into the eigenbasis, the susceptibility tensor can be expressed as summations of the product of a weight function and the particle-hole propagator over all possible quantum numbers, 
\begin{equation}\label{susr}
\chi_{ij}(\vec{r}) = \sum_{k_1 \lambda_{1}}\sum_{k_2 \lambda_{2}}
W_{ij}(\vec{r}) \:
\left[\frac{f(\epsilon_{k_{1}\lambda_{1}})-f(\epsilon_{k_{2}\lambda_{2}})}{\epsilon_{k_{2}\lambda_{2}}-\epsilon_{k_{1}\lambda_{1}}-i\eta}
\right],
\end{equation}
where $\epsilon_{k\lambda} = k^2/2m^*- \lambda k\gamma_R$ is the dispersion for the particle with momentum $k$ and chirality $\lambda$. The weight function inside the summations is
\begin{equation}
W_{ij}(\vec{r}) = 
e^{i(\vec{k_2}-\vec{k_1}) \cdot \vec{r}}
( u^{\dag}_{\lambda_1} \sigma^i u^{}_{\lambda_2}) ( u^{\dag}_{\lambda_2} \sigma^j u^{}_{\lambda_1}).
\end{equation}

While the derivation of the spin susceptibility tensor is straightforward, it still requires rather involved numerics. However, the numerical task can be greatly reduced by various symmetry arguments. Let's take the component $\chi_{xy}(\vec{r}) = \chi_{xy}(r,\theta)$ as a working example. Since the operators $\sigma_x, \sigma_y$ carry the 2D angular momentum $m = \pm 1$, making use of the SO(2) rotational invariance, the corresponding susceptibility $\chi_{xy}(r,\theta)$ contains linear combinations of $m=0, \pm 2$ by tensor analysis. That is to say, $\chi_{xy}(r,\theta) = f_0(r) + f_2(r) \cos 2\theta + g_2(r) \sin 2\theta$, where $f_0(r), f_2(r), g_2(r)$ are some functions without angular dependence. Furthermore, applying the parity symmetry in $y$ direction, it requires $\chi_{xy}(r,\theta) = -\chi_{xy}(r,-\theta)$ and enforces the functions $f_0(r), f_2(r)$ to vanish. Finally, the Onsager relation from the time-reversal symmetry indicates $\chi_{yx}(r,\theta) = \chi_{xy}(r,\theta+\pi) = g_2(r) \sin 2\theta$. Utilizing the rotational SO(2), parity $P_y$ (or equivalently $P_x$), and time reversal symmetries, one can work out the remaining components of the susceptibility tensor,
\begin{eqnarray}
\chi_{ij}(r,\theta)=\left[
\begin{array}{ccc}
g_0+g_2\cos2\theta & g_2\sin2\theta & g_1\cos\theta \\
g_2\sin2\theta & g_0-g_2\cos2\theta & g_1\sin\theta \\
-g_1\cos\theta & -g_1\sin\theta & h_0 \end{array} \right].
\end{eqnarray}
It is rather remarkable that the symmetry arguments make the angular dependence explicit and reduce the numerical task down to evaluation of four real scalar functions, $g_0(r)$, $g_1(r)$, $g_2(r)$ and $h_0(r)$. The Rashba Hamiltonian we study here further constrains $h_0(r) = g_0(r)+g_2(r)$, which reduces the number down to three.

\begin{figure}
\centering
\includegraphics[width=6cm]{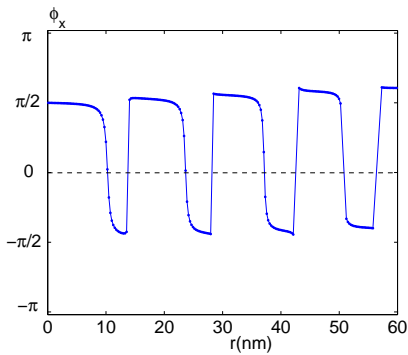}
\caption{\label{rkky}
(Color online) The spiral angle in weak Rashba regime with $k_R/k_F = 0.042$ at
$T=30$ K. The smooth minus $\pi$-jumps originate from the opposite
tendency of angle evolution of the Rashba spiral and RKKY effect.
}
\end{figure}

Suppose the ferromagnet on the left of the TMJ is aligned along the $z$-axis, we are interested in the mediated non-collinear exchange coupling proportional to $\chi_{iz}(r,\theta=0)$, where $r$ is the width of intermediate layer. Since $\chi_{yz}(r,0)=0$, the induced moment is captured by the spiral angle (shown in Fig.~\ref{TMJ}),
\begin{eqnarray}
\phi_x(r) = \tan^{-1} \left[ \frac{\chi_{zz}(r,0)}{\chi_{xz}(r,0)} \right] 
= \tan^{-1} \left[ \frac{g_0(r)+g_2(r)}{g_1(r)} \right].
\end{eqnarray}
For realistic materials\cite{Luo,Das,Nitta,Engle,Heida98}, we choose the spin splitting $\Delta_{R} \equiv 2 k_F \gamma_R = 5$ meV and the Fermi energy $\epsilon_F = 60$ meV. Or equivalently, it corresponds to the Rashba coupling $\gamma_R=8.91 \times 10^{-12}$ eV m and the carrier density $n_{2D} = 1.25 \times 10^{12}$ cm$^{-2}$. Thus, it falls into the weak Rashba regime, characterized by the dimensionless parameter $k_R/k_F = 0.042 \ll 1$, where $k_R=2 m^* \gamma_R$ and $k_F$ is the Fermi momentum in the absence of Rashba splitting. From Fig.~\ref{rkky}, it is clear that the non-collinear angle between the ferromagnets $\phi_x(r)$ shows RKKY-like oscillations with the gradual upswing trend due to the Rashba interaction. The numerical results, drastically different from the spin-precession argument in Datta-Das SFET, demonstrate the importance of the quantum interferences from all patches of the Fermi surface.

The RKKY oscillation with upward trend can be understood in a simple picture. Taking the asymptotic limit $k_F r \gg 1$, the reduced spin susceptibility along the radial direction $\chi_{ab}(r)$, where $a,b = x,z$, can be well approximated as 1D Rashba system. Applying a local gauge transformation, $U(r)=e^{-i k_R r \sigma^y/2}$\cite{Alei,Ima}, the Rashba Hamiltonian can be mapped into the 1D free electron gas with the well-known RKKY spin susceptibility. Since the local gauge transformation is nothing but the local rotation about the $y$-axis with the spiral angle $\phi(r) = k_R r$, the reduced susceptibility is approximately the usual RKKY oscillation twisted by a local spiral transformation,
\begin{equation}
\chi_{ab}(r) \approx \left[
\begin{array}{cc}
\cos k_R r & -\sin k_R r \\
\sin k_R r & \cos k_R r
\end{array} \right]_{ac} \chi^{RKKY}_{cb}(r),
\end{equation}
where the summation over the repeated index $c=x,z$ is implied. The gauge argument explains why our numerical results resemble the RKKY oscillation but with a gradual spiral background.

\begin{figure}
\centering
\includegraphics[width=6cm]{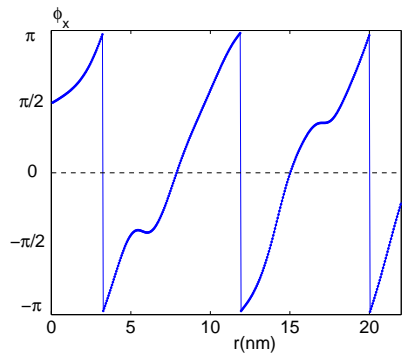}
\caption{\label{rashba}
(Color online)
The spiral angle in the strong Rashba regime where $k_R/k_F = 2.6$ and $T=30$ K. It is clear that the spiral exchange dominates with minor ripples from the RKKY interaction.}
\end{figure}

Note that the Fermi surface composed of the concentric circles with opposite chiralities is topologically equivalent to that for the 1D electron gas with opposite spins. It is exactly because of this topological equivalence so that the local gauge transformation is possible. Therefore, we were motivated to investigate what happens when the topology of the Fermi surface changes in the strong Rashba regime where only one chirality is present. By increasing Rashba coupling to $k_R/k_F =2.6$, our numerical results, shown in Fig.~\ref{rashba}, demonstrate robust spiral structure with minor oscillatory ripples. It is rather amazing that the change of Fermi surface topology swings the magnetic property from the RKKY dominated to the spiral. Another route to enter the strong Rashba regime is by reducing the carrier density. Since the transition occurs at $k_R/k_F =1$, for the Rashba coupling $\gamma_R=8.91 \times 10^{-12}$ eV m, one needs to reduce the density below $n_{2D} \sim 2.18 \times 10^9$ cm$^{-2}$, which may be achieved by applying gate voltage (see discussions below).

It is worth emphasizing that one should not confuse the robust spiral exchange here with the spin precession from single-particle argument. Since the chirality is even under time reversal transformation, the Kramers degeneracy connects opposite patches of the same circle. Therefore, the quantum interferences leading to RKKY oscillations are still present, as manifest in our numerical results. The puzzle is why the spiral interaction, when only one chirality is present, always takes the leading role, rendering the RKKY into minor ripples on the spiral backbone. At the point of writing, we do not have a simple physical interpretation for the interesting transition driven by the change of Fermi surface topology. But, as expected, the minor oscillatory ripples get further suppressed when the ratio $k_R/k_F$ increases.

\begin{figure}
\begin{center}
\includegraphics[width=6cm]{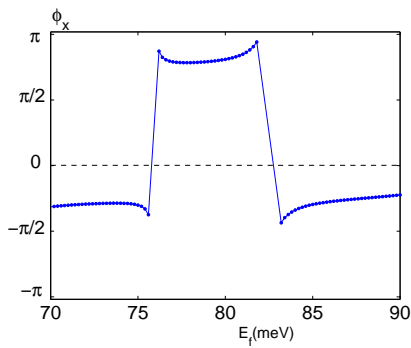}
\caption{\label{gate}The spiral angle versus the Fermi energy for the
junction of the width 90 nm at $T = 30$ K. The slightly upward tendency and
the non-symmetry location of each region is caused by the Rashba
spiral coupling.}
\end{center}
\end{figure}

In addition to its connection to Fermi surface topology, the mediated non-collinear coupling opens up potential applications in many magnetic devices by electrical manipulations. For instance, it was demonstrated in some experiments\cite{Nitta,Engle,Heida98} that the Rashba coupling can be controlled by the gate voltage. On phenomenology ground, the Rashba coupling reacts to the external electric field linearly, $\gamma_R = b \langle E\rangle$, where the coefficient $b$ is inversely proportional to the energy band gap and the effective mass\cite{Lommer}. Meanwhile, one can also use the external gate to manipulate the density of itinerant carriers. It comes into notice that the density concentration can be enhance up to 70\% (corresponding to 30 \% increase in Fermi energy) by electric means\cite{Nitta,Heida98}. Since the carrier density responds to external electric field more sensitively than the Rashba coupling, we numerically compute the non-collinear angle $\phi_x(r)$ at different Fermi energies but keeping the Rashba coupling $\gamma_R=8.91 \times 10^{-12}$ eV m fixed. As shown in Fig.~\ref{gate}, by changing the gate voltage, it is possible to induce sudden reversal of magnetic moments because of the RKKY oscillations. However, due to the presence of Rashba interaction, the angular jump deviates from $\pm \pi/2$ with upswing spiral background. Note that, before applying the idea of carrier-mediated non-collinear exchange coupling to realistic materials, one must keep in mind that the simple model we studied here does not include strain effects, disorders and surface roughness. In particular, the randomness of the interface can ruin our predictions and more careful ensemble average must be included.

While we mainly concentrate on the magnetic aspect of TMJ in previous paragraphs, the transport aspect is as important. In a recent paper\cite{Bauer03}, Bauer {\it et al.} showed the interesting universal angular magnetoresistance and also the spin torque in ferromagnetic/normal-metal heterostructures. Besides, Myers {\it et al.}\cite{Myers} demonstrated the possibility to reverse the domain orientation by the spin torque from the injected currents. Compared with the spin torque effect in magnetic tunneling junction\cite{Tulapurkar05}, the presence of the non-collinear exchange should deliver even richer phenomena. It is interesting to explore how the mediated non-collinear coupling and the spin torque compete and reshape our understanding in magnetic junctions.

In conclusion, we found the carrier-mediated non-collinear exchange coupling across the trilayer magnetic junction can not be explained by the simple spin-precession argument within single-particle picture. As the strength of the Rashba coupling increases, the mediated exchange goes from the oscillatory RKKY dominated to the robust spiral. Surprisingly, the change of magnetic behavior coincides with the transition of Fermi surface topology. As the nanotechnology advances in recent years, we believe that the clean and sharp interfaces can be realized in experiments and the effects we studied here would become important and measurable in experiments.

We acknowledges the grant supports from the National Science Council in Taiwan through NSC 94-2112-M-007-031(HHL), NSC 93-2112-M007-005 (HHL) and NSC 94-2112-M-009-025 (CHC). Financial supports for HHL through Ta-You Wu Fellow from National Center for Theoretical Sciences is also greatly appreciated.


\begin{thebibliography}{99}
\bibitem{Wolf01}
S. A. Wolf {\it et al.},
Science {\bf 294}, 1488 (2001).

\bibitem{Zutic04}
I. Zutic, J. Fabian and S. Das Sarma,
Re. Mod. Phys. {\bf 76}, 323 (2004).

\bibitem{MacDonald05}
A. H. MacDonald, P. Schiffer and N. Samarth,
Nature Mat. {\bf 4}, 195 (2005).

\bibitem{Sun04b}
S.-J. Sun and H.-H. Lin, Phys. Lett. A {\bf 327}, 73 (2004).

\bibitem{Sun06}
S.-J. Sun and H.-H. Lin, Eur. Phys. J. B {\bf 49}, 403 (2006).

\bibitem{Datta89}
S. Datta and B. Das, Appl. Phys. Lett. {\bf 56}, 665 (1989).

\bibitem{Murakami03}
S. Murakami, N. Nagaosa and S.-C. Zhang,
Science {\bf 301}, 1348 (2003).

\bibitem{Sinova04}
J. Sinova {\it et al.}, Phys. Rev. Lett. {\bf 92}, 126603 (2004).

\bibitem{Sun04a}
S.-J. Sun, S.-S. Cheng and H.-H. Lin, Appl. Phys. Lett.
{\bf 84}, 2862 (2004).
 
\bibitem{Luo}
J. Luo, H. Munekata, F.F. Fang and P.J. Stiles, Phys. Rev. B. {\bf
38}, R10142 (1988).

\bibitem{Das}
B. Das, D.C. Miller, S. Datta, R. Reifenberger, W.P. Hong, P.K.
Bhattacharya, J. Singh and M. Jaffe, Phys. Rev. B. {\bf 39},
1411 (1989).

\bibitem{Nitta}
J. Nitta, T. Akazaki, H. Takayanagi and T. Enoki, Phys. Rev. Lett.
{\bf 78}, 1335 (1997).

\bibitem{Engle}
G. Engels, J. Lange, Th. Sch\"{a}pers and H. L\"{u}th, Phys. Rev.
B. {\bf 55}, R1958 (1997).

\bibitem{Heida98}
J.P. Heida, B.J. van Wees, J.J. Kuipers, T.M. Klapwijk and G.
Borghs, Phys. Rev. B {\bf 57}, R11911(1998).

\bibitem{Alei}
I.L. Aleiner and V.I. Fal'ko, Phys. Rev. Lett. {\bf 87},
256801 (2001).

\bibitem{Ima}
H. Imamura, P. Bruno and Y. Utsumi, Phys. Rev. B {\bf 69},
121303(R) (2004).

\bibitem{Lommer}
G. Lommer, F. Malcher and U. R\"{o}ssler, Phys. Rev. Lett. {\bf
60}, 728 (1988).


%\bibitem{Yiping}
%Yiping. Lin, T. Koga and J. Hitta, Phys. Rev. B {\bf 71},
%045328(2005).


\bibitem{Bauer03}
G. E. W. Bauer, Y. Tserkovnyak, D. Huertas-Hernando and A. Brataas,
Phys. Rev. B {\bf 67}, 094421 (2003).

\bibitem{Myers}
E.B. Myers {\em et al.}, Science {\bf 285},867(1999).

\bibitem{Tulapurkar05}
A. A. Tulapurkar {\it et al.}, Nature {\bf 438}, 339 (2005).

%\bibitem{Loss}
%S.I. Erlingsson, J. Schliemann and D. Loss, Phys. Rev. B {\bf 71},
%035319(2005).

\end{thebibliography}
\end{document}